\documentclass[12pt]{iopart}
\usepackage{graphicx,color,psfrag}
\def\bc{\begin{center}}
\def\ec{\end{center}}

\def\beq{\begin{equation}}
\def\eeq{\end{equation}}

\begin{document}

\title{Valley symmetry breaking and gap tuning in graphene by spin doping }
\author{Antonio Hill, Andreas Sinner and Klaus Ziegler}
\address{Institut f\"ur Physik, Universit\"at Augsburg}
\date{\today}

\begin{abstract}
We study graphene with an adsorbed spin texture, where the localized spins create a periodic
magnetic flux. The latter produces gaps in the graphene spectrum and breaks the valley symmetry.
The resulting effective electronic model, which is similar to Haldane's periodic flux model,
allows us to tune the gap of one valley independently from that of the other valley. This leads
to the formation of two Hall plateaux and a quantum Hall transition. 
We discuss the density of states, optical longitudinal and Hall 
conductivities for nonzero frequencies and nonzero temperatures. 
A robust logarithmic singularity appears in the Hall conductivity when
the frequency of the external field agrees with the width of the gap.  
\end{abstract}

\maketitle

\section{Introduction}

Transport properties of neutral graphene are characterized by a semimetallic behavior with
a point-like Fermi surface at two valleys. This can be changed
by external electric fields in gated graphene to more classical transport with a circular
Fermi surface or by an external magnetic field that opens gaps in the spectrum
~\cite{novoselov04,novoselov06}. The latter
leads to the formation of Hall plateaux in the Hall conductivity. Another possibility is
to introduce a gap either by chemical doping (e.g. with hydrogen \cite{elias09}) or in bilayer
graphene with a double gate \cite{ohta06,castro07,min07,oostinga08,zhang09,mak09}. 
Chemical doping is an interesting direction for
modifying graphene because it leads to a rich field with new properties. Although a standard
technique for semiconductors, it has been applied to graphene only recently 
\cite{elias09,zhou07,gietz08,zhou08,mak09,bangert10,coletti10,mcchesney10}. 
As a special case of chemical doping one can use
atoms that carry a spin to create a spin texture on graphene. These localized spins alter
the transport properties of graphene significantly. This will be discussed in this paper.

The quantum Hall effect in graphene was observed in a number of experiments, 
cf. Ref~\cite{novoselov04,novoselov06}. It is usually associated with the presence of 
an external homogeneous magnetic field which separates the electronic spectrum into Landau levels. 
Without an external magnetic field each valley of the honeycomb spectrum provides a quantum Hall step
\cite{semenoff84}. However, the system is time-reversal invariant. This implies that the Hall
conductivity vanishes, since the contribution of the two valleys 
to the Hall conductivity cancel each other. A magnetic flux, however, can cure this 
problem by breaking the time-reversal invariance. 
A possible way to observe the quantum Hall effect is by introducing a periodic magnetic flux.
This was suggested by Haldane \cite{Haldane88a}: 
A staggered magnetic flux, in combination with nearest and next-nearest neighbor hopping and a staggered 
potential on the honeycomb lattice, creates two Hall plateaux with Hall conductivities 
$\sigma_{xy}=\pm e^2/h$.
The periodic magnetic field leads to a staggered flux 
with zero net flux in each unit cell. It affects the hopping matrix elements by 
creating a phase factor which enables us to change the signs of the Hall conductivities of each 
node independently by changing the magnetic field strength. 
The effects of the staggered potential and the staggered magnetic flux can also be understood 
in terms of symmetry breaking: The Brillouin zone has a six-fold energetic degeneracy due to
the vanishing energy at the corners of its hexagonal structure. A staggered potential breaks the
inversion symmetry \cite{semenoff84,Haldane88a} but preserves the six-fold energetic 
degeneracy, because all six corners acquire the same gap. The staggered magnetic flux, however,
reduces the six-fold degeneracy to a three-fold degeneracy (i.e. it provides inter-valley 
symmetry breaking) because it affects the gap of the two Dirac nodes differently.
 
The importance of sublattice and inter-valley symmetry breaking in graphene for applications
has been widely recognized. Recently the effect of the broken inversion symmetry in graphene
was addressed to in context of topological \cite{Niu07} and anomalous thermoelectric
\cite{DSarma09} transport, as well as valley-dependent optoelectronics \cite{Niu08}. 

\begin{figure}[t]
\centering
\includegraphics[width=5cm]{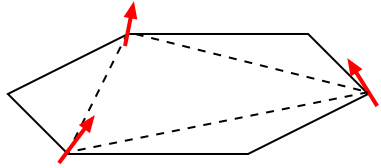}
\caption{Spinor atoms on a unit cell of graphene. Tilted spin textures can provide an effective 
staggered magnetic 
flux~\cite{mueller-h96,ohgishi00,campana06}.}
\label{fig:Spins}
\end{figure}

In this paper we propose a model that is based on a honeycomb lattice, where one sublattice
is occupied by localized spins. This leads to an effective hopping parameter with a Berry
phase and reflects the equivalence of the spin texture with an effective periodic magnetic
flux in the tight-binding model. It is similar but slightly different from Haldane's periodic
flux model.
The inter-valley symmetry is lifted by introducing only a next-nearest neighbor hopping 
on one sublattice and a staggered 
magnetic flux. This choice is motivated by the fact that graphene on a substrate can adsorb atoms 
only on one side. In this case the adsorbed atoms occupy only next-nearest neighbor sites on 
the honeycomb lattice, i.e. they occupy one sublattice.
Such a possibility was experimentally observed for hydrogen on graphene \cite{elias09}. 
As a result, the adsorbed atoms modify the overlap integrals on their sublattice and
may even create an additional next-nearest neighbor hopping. Moreover, if the adsorbed
atoms have a magnetic moment, these moments can form a localized spin texture due to double 
exchange~\cite{zener51,anderson55,desgennes60}. It is known that
such spin structures provide an effective Berry phase for the electron 
hopping rate~\cite{mueller-h96,ohgishi00,campana06}. 
The latter has the same effect as a periodic (staggered) flux,
where the flux depends on the tilting angle of the localized spins (Figs.~\ref{fig:Spins},
\ref{fig:texture}). 

\section{Model: electronic hopping on a spin texture}

We consider localized $S=1/2$ spins on a graphene lattice, where the latter is assumed 
to be flat. As a reference system we choose a basis
of eigenspins oriented in $z$-direction with the lattice in the $x$-$y$ plane. 
In other words, the $z$ component of the localized spin 1/2 operator satisfies 
$S^z|\pm\rangle=\pm|\pm\rangle$. In terms
of Pauli matrices ${\vec \sigma}=(\sigma_x,\sigma_y,\sigma_z)$ this reads
\beq
\sigma_z {\bf s}_\pm=\pm {\bf s}_\pm 
\label{eigenspin}
\eeq
with the two-component spinor basis $\{{\bf s}_+,{\bf s}_-\}$. In general, a spin state
can have a local orientation, such that it is an eigenstate to 
${\vec n}_j\cdot{\vec \sigma}$ with the 3D unit vector 
${\vec n}_j=(\sin\theta\cos\phi,\sin\theta\sin\phi,\cos\theta)$. The matrix of the spin operator 
${\vec n}_j\cdot{\vec \sigma}$ reads in the $z$-oriented basis
\[
{\vec n}_j\cdot{\vec \sigma}=\pmatrix{
\cos\theta_j & e^{i\phi_j}\sin\theta_j \cr
e^{i\phi_j}\sin\theta_j & -\cos\theta_j \cr
}
\]
and 
\[
{\bf s}_j={\vec n}_j\cdot{\vec \sigma}{\bf s}_+
=e^{ib_j}\pmatrix{
\cos\theta_j  \cr
e^{i\phi_j}\sin \theta_j  \cr
}  
\ .
\]
Here the tilting angle $\theta_j$ and the $x$-$y$ rotation angle $\phi_j$ 
refer to the change of the quantization direction, relative to the spin state 
${\bf s}_+$.

\begin{figure}[t]
\centering
\includegraphics[width=8cm]{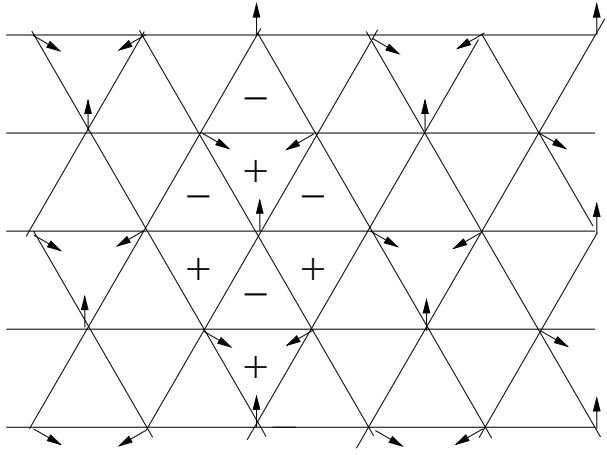}
\caption{Spin texture on the triangular sublattice with fixed tilting angles 
(all $x$-$y$ projected spins have the same length) and $x$-$y$ rotation angels with multiples of $2\pi/3$.
The + (-) signs indicate a positive (negative) magnetic flux through the corresponding
(up or down) triangle, whose value is determined by Eq. (\ref{peierls}).}
\label{fig:texture}
\end{figure}

Then the electronic Hamiltonian on a honeycomb lattice must describe the
interaction of the electronic spin with the localized spins. This leads to 
a spin-dependent hopping amplitude~\cite{mueller-h96,ohgishi00,campana06}:
\beq
{\cal H}=\sum_{<i,j>}H_{ij}c^\dagger_ic_j+h.c.
\label{hamiltonian00}
\eeq
with 
\beq
H_{ij}=t{\bf s}_i\cdot {\bf s}_j =te^{ia_{ij}}\cos(\theta_{ij}/2)
\label{hopping1}
\eeq
and electronic creation (annihilation) operators $c^\dagger_i$ ($c_j$).
The Berry phase $a_{ij}$ is given as~\cite{campana06}:
\beq
\sin a_{ij}=-\frac{\sin(\theta_{i}/2)\sin(\theta_j/2)\sin(\phi_i-\phi_j)}{\cos(\theta_{ij}/2)}
\label{peierls}
\eeq
which implies $a_{ij}=-a_{ji}$. (It should be noticed that the relation (\ref{peierls}) means
that $a_{ij}$ is the solid angle spanned by ${\bf s}_i,{\bf s}_j$ 
and the $z$ axis~\cite{campana06} .)
The phase $a_{ij}$ vanishes for vanishing tilting angles $\theta_i$,
$\theta_j$ (i.e. for ferro- or antiferromagnetic states)
and for $\phi_i=\phi_j$ (i.e. when the spin projections on the $x$-$y$ plane are parallel.
Eq. (\ref{hopping1}) gives us the general expression for the hopping of the electrons in graphene
with an additional spin texture, where the latter is characterized by the tilting angles and 
$x$-$y$ rotation angles. If we assume
a specific configuration of localized spins, the corresponding quantization axis of the spin 
${\bf s}_j$ is fixed by the angles $\theta_j$ and $\phi_j$. A special case, where all
angles $\theta_{ij}$ between next-nearest neighbor spins and all tilting angles $\theta_j$
are equal (except for an irrelevant sign change of $\theta_{ij}$ for different pairs $i,j$)
is depicted in Fig. \ref{fig:texture}.
This is similar to the situation on the Kagom\'e lattice in Ref.~\cite{campana06}.  
Moreover, the angles $\phi_j$ are rotating by $2\pi/3$ on the lattice such that their differences
in the expression of the phase $a_{ij}$ in Eq. (\ref{peierls}) are $\pm 2\pi/3$ for next nearest
neighbors. This implies a global renormalization of the hopping rate $t\to t\cos\theta_{ij}$
and a Berry phase that changes only its sign from lattice bond to lattice bond. 

\subsection{Tight-binding model on the honeycomb lattice}

We begin with the tight-binding Hamiltonian of monolayer graphene (i.e. for
a honeycomb lattice) and return to
the effect of the spin texture later. Then in Fourier space we can write
for the Hamiltonian matrix in sublattice representation
\begin{equation} 
H = h^{}_1\sigma^{}_1 + h^{}_2\sigma^{}_2 + h^{}_3\sigma^{}_3 \ ,
\label{eq:model}
\end{equation}
where the off-diagonal Pauli matrices describe the hopping between the two 
sublattices. The diagonal Pauli matrix term with $h^{}_3$ describes processes
on the same sublattice that can include next-nearest neighbor tunneling.
The eigenvalues of the Hamiltonian are 
\begin{equation}
\label{eq:spectrum} 
E^{}_\pm=\pm E^{}_k = \pm\sqrt{h^2_1+h^2_2+h^2_3}.
\end{equation}
Specifically for the honeycomb lattice we have for the nearest-neighbor terms
\begin{equation} 
\label{eq:h1}
h^{}_{1} = -t\sum_{i=1}^{3}\cos({\bf a}^{}_i\cdot {\bf k}) , \ \ \
h^{}_{2} = -t\sum_{i=1}^{3}\sin({\bf a}^{}_i\cdot {\bf k})
\label{h_12}
\end{equation}
with the basis vectors of the honeycomb lattice ${\bf a}^{}_i$ 
given by
\begin{equation} 
\label{eq:va1}
{\bf a}^{}_1 = a\left(0,-1\right), \ \ \
{\bf a}^{}_{2,3} = \displaystyle \frac{a}{2}\left(\pm\sqrt{3},1\right),
\end{equation}
where $a$ denotes the lattice constant. 
In absence of a gap, i.e. for $h^{}_3 = 0$, the spectrum  defined in Eq.~(\ref{eq:spectrum}) 
vanishes at nodal points whose positions in the Fourier space are given by the vectors
\beq
{\bf b}^{\pm}_1 = \frac{4\pi}{3\sqrt{3}a}\left(\pm 1, 0\right),\ \ \
{\bf b}^{\pm}_2 = \frac{2\pi}{3\sqrt{3}a}\left(-1,\pm\sqrt{3}\right),\ \ \
{\bf b}^{\pm}_3 = \frac{2\pi}{3\sqrt{3}a}\left(1,\pm\sqrt{3}\right).
\eeq
Assuming a uniform gap, i.e. $h^{}_3=m$, the Hamiltonian defined in Eq.~(\ref{eq:model}) 
can be approximated in the vicinity of the nodal points by the Hamiltonian of a massive Dirac 
fermion
\begin{equation}
\label{eq:LowEnModel} 
H \approx vk^{}_1\sigma^{}_1 + vk^{}_2\sigma^{}_2 + m\sigma^{}_3,
\end{equation}
where $v=\sqrt{3}ta/2=1$ denotes the Fermi velocity of electrons. With this
the eigenvalues read 
\begin{equation}
\label{eq:EVleh} 
E^{}_{\pm} = \pm E^{}_k \approx\pm\sqrt{k^2+m^2}
\ .
\end{equation}

\begin{figure}[t]
\centering
\includegraphics[width=5cm]{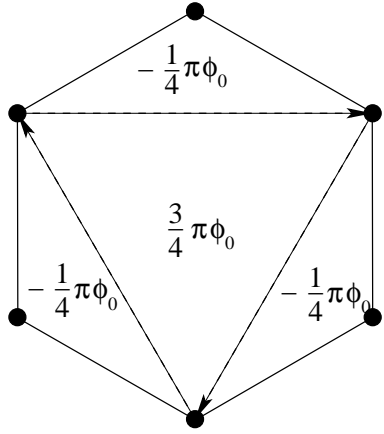}
\caption{Magnetic flux through the graphene unit cell corresponding to the potential 
Eq.~(\ref{eq:FirstQuant}). Arrows indicate second-neighbor hopping processes and $\phi_0=h/e$ is the
flux quantum.}
\label{fig:Flux}
\end{figure}

\begin{figure}[t]
\centering
\includegraphics[width=9cm]{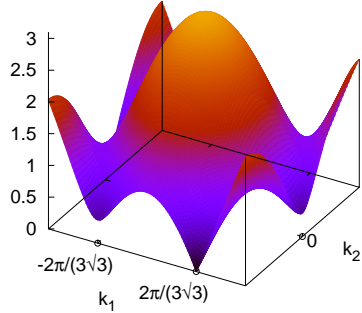}
\caption{Upper band of the full spectrum of the tight-binding Hamiltonian with the 
nonuniform gap defined in Eqs.~(\ref{h_12}), (\ref{h3}).}
\label{fig:Spectrum3D}
\end{figure}

Now we include the spin texture, described by the modified hopping term of Eq. (\ref{hopping1}),
by the phase variable $\phi_{{\bf r},{\bf r}^\prime}$ which is the Berry phase of Eq. (\ref{peierls}). 
Using a spin texture as shown in Fig. \ref{fig:texture}, we get 
a Berry phase $\phi$ which is the same for all three next-nearest neighbor vectors ${\bf c}_j$ ($j=1,2,3$
(cf. Fig. \ref{fig:Flux})
\begin{equation}
\label{eq:cv1} 
{\bf c}^{}_1 = \sqrt{3}a\left(1,0\right), \ \ \
{\bf c}^{}_{2,3} = \displaystyle \frac{\sqrt{3}a}{2}\left(-1,\pm\sqrt{3}\right)
\ ,
\end{equation}
because the tilting angle is fixed and the $x$-$y$ rotation angle of the spins is 
$\pm 2\pi/3$ between next-nearest neighbor sites. Then the next-nearest neighbor hopping term reads
\beq
\chi^{}_{{\bf r},{\bf r}^\prime}=t'\sum^{3}_{j=1}
\left( 
e^{-i\phi}
\delta^{}_{{\bf r},{\bf r}^\prime-{\bf c}^{}_j} +
e^{i\phi}
\delta^{}_{{\bf r},{\bf r}^\prime+{\bf c}^{}_j}
\right)
\  .
\label{NNN}
\eeq
$t'$ is the renormalized hopping parameter of Eq. (\ref{hopping1}). 
This hopping term is combined with a uniform gap to give in real space with sublattice
co-ordinates ${\bf r}$
\beq
\label{eq:FirstQuant} 
h^{}_{3;{\bf r},{\bf r}^\prime} = M~\delta^{}_{{\bf r},{\bf r}^\prime} 
+\chi^{}_{{\bf r},{\bf r}^\prime}
\label{h_3}
\eeq
for the third term in Eq. (\ref{eq:model}). For the diagonal term $M$ represents a potential
contribution of the doping atoms on the sublattice (i.e. there is a potential difference of $2M$
between the two sublattices) and $t'$ is the contribution of the doping atoms on the next-nearest
neighbor hopping term. After Fourier transformation we get for $h_3$
\beq
\label{eq:shift1}
h_3=M+\chi^{}_{\bf k} 
=M+2t'\sum_{i=1}^3 \cos\left({\bf c}^{}_i\cdot{\bf k}-\phi\right)
\ .
\label{h3}
\eeq
The term $h_3\sigma_3$ in $H$ of Eq. (\ref{eq:model}) opens a gap in the dispersion
$\sqrt{h_1^2+h_2^2+h_3^2}$ if it is nonzero at the nodal points (valleys) of 
$\sqrt{h_1^2+h_2^2}$. Since $h_3$ is also ${\bf k}$ dependent it can vary the gap
independently for the two valleys. It can be adjusted such that the gap parameters
have different signs or the gap vanishes 
at one valley but provides a gap at the other valley.
For instance, with $\phi=-\pi/4$ and $M=3t'(1+\sqrt{3})/\sqrt{2}$ this term 
vanishes at the nodal points ${\bf b}^{+}_1$ and ${\bf b}^{\pm}_2$ and
remains nonzero at the nodal points ${\bf b}^{-}_1$ and ${\bf b}^{\pm}_3$. 
The magnetic flux through a unit cell is shown in Fig. \ref{fig:Flux} and the
related spectrum is presented  in Fig.~\ref{fig:Spectrum3D}.

This model is similar to the periodic flux model introduced by Haldane \cite{Haldane88a}.
However, the spin texture leads to some differences. 
In Table~\ref{compare} we compare the gap-opening term $h_3$ of the spin-texture model with 
the corresponding term in Haldane's model.

\begin{table}
\begin{center}
\begin{tabular}{ccc}
 & Haldane's model & spin texture model \\
   &  &  \\
$h_3$:\ \  & $M-\chi{'}_{\bf k}$\ \  & $M+\chi^{}_{\bf k}$ \\
  &  &  \\
$\chi{'}_{\bf k}$:\ \  & $2t^{}_2\sin\phi\sum_{j}^{AB}\sin({\bf k}\cdot{\bf c}^{}_j)$\ \  & $2t'\sum_{j}^A
\cos({\bf k}\cdot{\bf c}^{}_j-\phi)$ \\
\end{tabular}
\caption[smallcaption]{Comparison of Haldane's model with our model in terms of the
symmetry-breaking quantity $h_3$. The vector ${\bf c}_j$ connects next-nearest neighbor sites.
The summation runs either over both sublattices ($\sum_{j}^{AB}$) or only over sublattice A ($\sum_{j}^{A}$).
The definition of the variables of Haldane's model are those of Ref. \cite{Haldane88a}.}
\label{compare}
\end{center}
\end{table}
Below we will examine how the broken valley symmetry affects the transport properties.
For this purpose we focus on $\phi=-\pi/4$ and compare the results with gapless graphene
($M=t'=0$) and with graphene with uniform gap ($M\ne0$, $t'=0$).

\section{Density of states}

The density of states (DOS) is the imaginary part of the single particle Green function
\begin{equation}
\label{eq:DOS}
\rho(E) = -{\rm Im}~
{\rm Tr}_2~\frac{1}{\pi}
\int_{BZ}
~\Big(H-E+i0^{+}\Big)^{-1} \frac{d^2k}{\Omega^{}_{BZ}}
.
\end{equation}
Here, the operator ${\rm Tr}_2$ acts on the pseudospinor space and the momentum integral is taken over the 
Brillouin zone (BZ) of the honeycomb lattice, whereas $\Omega^{}_{BZ}$ denotes its volume. 

For the low-energy model of Eq.~(\ref{eq:LowEnModel}) the DOS can be evaluated explicitly when 
the integration over the Brillouin zone is replaced by one over a circular area with radius $\lambda$:
\begin{equation}
\nonumber
\rho(E) = {\rm Im}\intop_0^\lambda
\left( 
\frac{1}{E-E^{}_k-i0^{+}}+\frac{1}{E+E^{}_k-i0^{+}}
\right)k\frac{dk}{2\pi^2}
\end{equation}
which gives after integration in the limit $\lambda\to\infty$
\begin{equation}
\label{eq:DODres}
\rho(E) = \frac{E}{2\pi}\left[\Theta(E-|m|)-\Theta(-E-|m|)\right]
\ .
\end{equation}
This means that there are no states with nonzero energy for $-|m|\le E\le |m|$, where $m$ is
the effective gap determined by the value of $h_3$ at the gapped valley. 
For the case of gapless graphene, i.e. for $M=t'=0$,  Eq.~(\ref{eq:DODres}) reduces to
\begin{equation}
\label{eq:DODcl}
\rho(E) = \frac{E}{2\pi}\left[\Theta(E)-\Theta(-E)\right] = \frac{|E|}{2\pi}.
\end{equation}
The evaluation of the DOS for the model with the full spectrum and both uniform and nonuniform gap 
(i.e. with $|h^{}_3|=m$) can be performed numerically, since the integration has to be 
carried out over the whole Brillouin zone.
\begin{figure}[t]
\centering
\includegraphics[width=8cm]{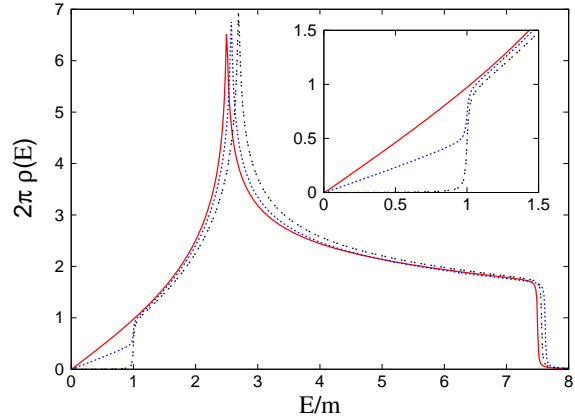}
\caption{Density of states of the full tight-binding model is evaluated.
Solid (red) line corresponds to gapless graphene; dashed (black) line is the DOS of the model with uniform gap 
($|h^{}_3|=m=0.4t$, $t'=0$); dotted (blue) line represents the DOS of the model for nonuniform gap ($t'\ne0$) 
with the same $m$. 
The inset shows the same curves for small energies. The van Hove singularity for gapless graphene 
lies at the hopping energy $E=t$. The shifting of the positions of the van Hove singularity 
as well as the change of the slopes at low energies should be
noticed upon a change of the parameters $m$ and $t'$.}
\label{fig:DOS}
\end{figure}
The DOS of graphene for different gap realizations is shown in Fig.~\ref{fig:DOS}. 
For small energies the DOS of gapless graphene shown as a (red) solid curve exhibits a linear behavior 
of the single Dirac cone model predicted by Eq.~(\ref{eq:DODcl}) (note the inset in Fig.~\ref{fig:DOS}). 
The DOS of the gapped model shown as (black) dashed line is zero for energies smaller than the gap $m$ 
and experiences a jump from zero to a finite value ($|m|/2\pi$ in the linear limit) at the energy $E=m$. 
The DOS of the model with the nonuniform gap depicted as (blue) dotted line in Fig.~\ref{fig:DOS} 
demonstrates an intermediate behavior in comparison to the other models. On the one hand it behaves 
linearly for energies smaller than the gap $m$. But the slope 
of the linear asymptote is only half of the corresponding value of gapless graphene, 
since the number of preserved Dirac cones in the Brillouin zone is half the number of Dirac cones of 
the gapless model. On the other hand DOS also reveals a discontinuous behavior at $E=m$ where the 
energy becomes sufficient to allow electrons to tunnel through the potential barrier of $2m$ that 
separates valence band from the conductivity band. The position of van Hove singularity related peaks 
in the DOS reveals dependence on the gap $m$ and for $m=0$ lies at the next-neighbor hopping energy 
$E=t$ as it can be seen from Fig.~\ref{fig:DOS}. 

\section{Optical conductivities}

Within the linear response approach the conductivity tensor for Hamiltonian $H$ is given by the Kubo formula.
Then we obtain for the conductivity at inverse temperature $\beta$ \cite{t2:ziegler06e}
\[
\sigma_{\mu \nu}(\omega) = \frac{i}{\hbar} \lim_{\alpha \rightarrow 0} \int_{BZ}
\sum_{l,l'=0,1}\, \langle k,l | j_\mu | k,l' \rangle \langle k,l'| j_\nu |k,l\rangle 
\]
\beq
\times\frac{1}{E_{k,l} - E_{k,l'}} 
\frac{f_\beta(E_{k,l'}) - f_\beta(E_{k,l})}{E_{k,l} - E_{k,l'} + \omega - i\alpha} \ \frac{d^2 k}{\Omega^{}_{BZ}} \, ,
\label{kubo-formel}
\end{equation}
where $f_\beta(E)=1/(1+\exp(\beta (E-E_F)))$ the Fermi-Dirac distribution at the inverse temperature $\beta$, 
$E_F$ is the Fermi energy, and $\omega$ is the frequency of the external field. $| k,l \rangle$
is eigenstate of $H$ with eigenvalue $E_{k,l}=(-1)^lE_k$. The index $l$ 
refers to the upper ($l=0$) and lower ($l=1$) band, respectively. Moreover, the current operator reads
\[
j_\mu =ie [H,r_\mu] 
\ .
 \] 
For the off-diagonal matrix elements with $l'\ne l$ we obtain from the Hamiltonian in Eq.~(\ref{eq:model}) 
\beq
\langle k,l | j_\mu | k,l' \rangle \langle k,l' | j_\nu | k,l \rangle 
=\frac{p_{\mu,a} p_{\nu,b}}{E^2_k}\left( E^2_k \delta^{}_{ab} -  h^{}_a h^{}_b\right)
+(-1)^l i\epsilon^{}_{abc}\frac{p_{\mu,a} p_{\nu,b} h_c}{E^{}_k},
\label{ctensor0}
\eeq
\[
p_{\nu,a}=\frac{\partial h_{a}}{\partial k^{}_\nu}
\ .
\]
The low-energy (Dirac) Hamiltonian, defined in Eq.~(\ref{eq:LowEnModel}) with $h_a=k_a$ ($a=1,2$)
and $h_3=m$, gives the following expression 
for the current tensor:
\beq
\langle k,l | j_\mu | k,l' \rangle \langle k,l' | j_\nu | k,l \rangle 
=e^2\left[
\delta_{\mu\nu}-\frac{k_\mu k_\nu}{E_k^2}+(-1)^l i\epsilon_{\mu\nu 3}\frac{m}{E_k}
\right]
\ .
\label{hallcond3}
\eeq
In case of the low-energy approximation the integration over the Brillouin zone can be 
approximated by a circular area with radius $\lambda$.
The longitudinal conductivity ($\mu=\nu$) is real due to the diagonal expression in Eq. (\ref{hallcond3}).
On the other hand, for the Hall conductivity, where $\mu\ne \nu$, the matrix element is complex:
\[
\langle k,l | j_\mu | k,l' \rangle \langle k,l' | j_\mu | k,l \rangle 
=e^2\frac{E^2_k-k^2_\mu}{E^2_k} 
, 
\]
\beq
\langle k,l | j_\mu | k,l' \rangle \langle k,l' | j_\nu | k,l \rangle 
= -e^2\frac{k^{}_\mu k^{}_\nu -(-1)^l i\epsilon^{}_{\mu\nu3} m E^{}_k}{E^2_k}.
\label{kubo_tensors}
\eeq
From the Kubo formula of Eq.~(\ref{kubo-formel}) we then obtain for the real part of the conductivity
\[
\sigma^{\prime}_{\mu\mu}(\omega) =
\]
\beq
 \frac{e^2}{\hbar}
\intop_0^\lambda~\frac{E^2_k+m^2}{8 E^3_k}~[f^{}_\beta(E^{}_k)-f^{}_\beta(-E^{}_k)]
\left[\delta(\omega-2E^{}_k) + \delta(\omega+2E^{}_k)\right] kdk
\ .
\label{eq:ZwErg}
\eeq
Finally, taking spin and valley degeneracy into account (factor 4) and sending $\lambda\to\infty$ 
we obtain~\cite{gusynin07,t2:ziegler07e}
\[
\sigma^{\prime}_{\mu\mu}(\omega) =
\]
\beq
\frac{\pi e^2}{2h}\left(1+\frac{(2m)^2}{\omega^2}\right)~[f^{}_\beta(\omega/2)-f^{}_\beta(-\omega/2)] 
\left[\Theta(\omega-2m)+\Theta(-\omega-2m)\right],
\label{eq:LongCond}
\eeq
which can be expressed as function of three dimensionless variables as
\begin{equation}
\sigma^{\prime}_{\mu\mu}(\omega) =  \frac{\pi e^2}{2h} f(\beta m,\beta\mu,\beta \omega).
\end{equation}
Hence the longitudinal optical conductivity vanishes for $|\omega|<2|m|$ and is nonzero for $|\omega|\ge 2|m|$. 
For zero temperature the Fermi functions can be replaced by Heaviside step functions 
$\lim_{\beta\to\infty} f^{}_\beta(\pm \omega) = \Theta(\pm \omega)$
and we obtain for the $ac$-conductivity the well-known result~\cite{gusynin07,t2:ziegler07e}
\begin{equation}
\sigma^{\prime}_{\mu\mu} (\omega)= \frac{\pi e^2}{2h} 
\end{equation}
for frequencies within the electronic bands.
This value has been found to be very robust and barely dependent on the temperature or quality of the 
sample.

One of the motivations for this work was to study the Hall conductivity in case of a broken valley symmetry, 
where we have a gapped and a gapless valley. Using the off-diagonal current tensor of Eq.~(\ref{kubo_tensors}),
the integration over the Brillouin zone is approximated again by the low-energy Hamiltonian. 
For a uniform gap (i.e. for $t'=0$)
the rotational symmetry of the problem is restored and therefore the contribution of $k_\mu k_\nu$ in the
off-diagonal current tensor vanishes. Thus, after sending the cut-off $\lambda\to\infty$, we get 
\[
\sigma^{}_{\mu\nu}(\omega)=-\epsilon^{}_{\mu\nu3}\frac{e^2}{\hbar}~\frac{m}{4\pi}
\intop^{\infty}_{|m|}\left[f^{}_\beta(E)-f^{}_\beta(-E)\right]
\]
\beq
\times\left(\frac{1}{\omega-2E-i0^{+}}-\frac{1}{\omega+2E-i0^{+}}\right)\frac{dE}{E} .
\label{eq:Sig_12Mast}
\eeq
For $|E^{}_F|\le |m|$ and $T=0$ the integral gives
\begin{equation}
\label{eq:ReSig_12} 
\sigma^{\prime}_{\mu\nu}(\omega) = \epsilon^{}_{\mu\nu3}~\frac{m}{4\pi\omega}
\log\left|\frac{2m+\omega}{2m-\omega}\right|. 
\end{equation}
At frequency $\omega=2m$ there is a logarithmic divergence in the Hall conductivity. For large frequencies 
$\omega$ this expression approaches zero, while for any nonzero gap and $\omega\to0$ it approaches 
the constant value that depends only on the sign of $m$
\begin{equation}
\sigma^{\prime}_{12} \approx \frac{m}{4\pi|m|}\frac{e^2}{\hbar}
= \frac{{\rm sgn}(m)}{2}\frac{e^2}{h}
\ .
\label{eq:hallcond1}
\end{equation}
These are the well-known Hall plateaux of Dirac fermions $\pm e^2/h$~\cite{semenoff84}. 
This result implies for two valleys with gap parameters $m$ and $m'$  \cite{Haldane88a}
\beq
\sigma^{\prime}_{12} \approx\left[
{\rm sgn}(m)-{\rm sgn}(m')\right]\frac{e^2}{2h}
\ .
\label{fullHall}
\eeq

\begin{figure}[t]
\centering
\includegraphics[width=8cm]{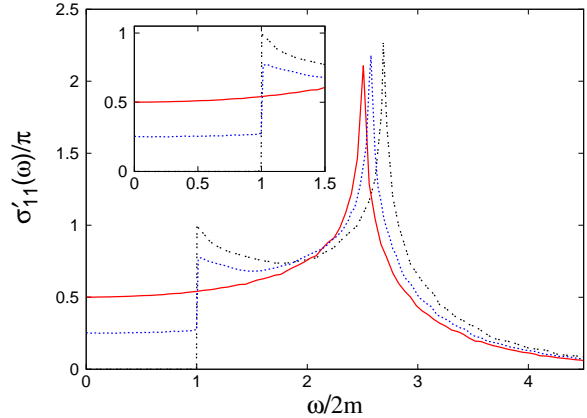}
\caption{Real part of the longitudinal optical conductivity at half-filling for $T=0 K$ 
and $m=0.4t$ in units of $e^2/h$. Here we show optical conductivities for: 
a) gapless graphene (red solid line); b) model with uniform gap (black dashed line); 
c) model with nonuniform gap (blue dotted line). The inset shows the same curves for 
small frequencies.  Positions of the van Hove related peaks coincide with those of DOS 
if one relates frequencies to the energies from Fig.~\ref{fig:DOS} by $\omega=2E$.}
\label{fig:long1}
\end{figure}

\begin{figure}[t]
\centering
\includegraphics[width=8cm]{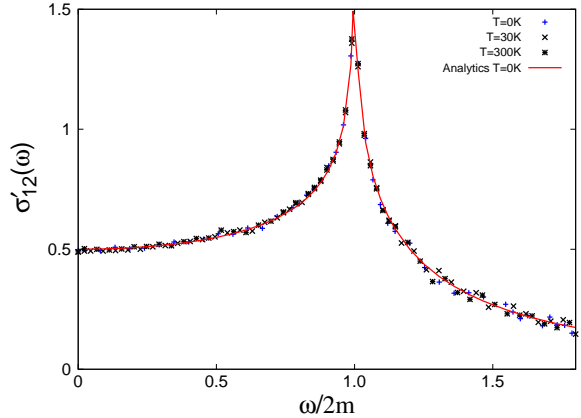}
\caption{Real part of the optical Hall conductivity at half-filling for $T= 0K,30K$ and $300K$ 
in units of $e^2/h$. Numerical results calculated for the full tight-binding model (dots and crosses) and
compared with the low-energy approximation of Eq.~(\ref{eq:ReSig_12}) (red solid line).}
\label{fig:Hall}
\end{figure}

\section{Discussion and Conclusions}

An evaluation of the DOS (Eq.~(\ref{eq:DOS})), the longitudinal conductivity 
and the Hall conductivity (Eqs.~(\ref{kubo-formel}), (\ref{ctensor0})) are performed numerically for 
the full tight-binding spectrum.
The results are presented in Figs.~\ref{fig:DOS}-\ref{fig:Hall} 
as a function of energy (DOS) or as functions of frequency (conductivities)
for nonzero temperatures. We compare the situation without spin texture ($t'=0$), both for the
gapless case $m=0$ and the uniform gap $m\ne 0$ and the situation with spin texture ($t'\ne0$).
In the latter we fix the Berry phase by $\phi=-\pi/4$ and the uniform
gap parameter by $M=3t'(1+\sqrt{3})/\sqrt{2}$
and call this the asymmetric valley, whereas the case $t'=0$ is the symmetric valley.

{\it DOS}: The effect of the periodic magnetic flux, as described by $h_3$ in Eq. (\ref{h3}),
is reflected by the DOS in Fig. \ref{fig:DOS}. There is either a full gap (symmetric valleys) 
or a gapless DOS (asymmetric valley). The DOS of the asymmetric valley has a reduced slope 
in comparison with the case where both valleys are gapless ($M=t'=0$), 
since only one valley contributes at low energies.
The position of the van Hove singularity depends on the parameters $M$, $t'$ and moves to higher
energies as we switch on $M$ and $t'$.

{\it longitudinal conductivity}:
The behavior of the longitudinal optical conductivity in Fig. \ref{fig:long1} for $M=t'=0$
reproduces the constant universal value $\sigma_{\mu\mu}'=\pi e^2/2h$ at low frequencies.
At higher frequencies it increases due to the van Hove singularity, just like the DOS.
This is also the case for $M\ne 0$, where the gap for $t'=0$ leads to a vanishing conductivity
for frequencies less than the gap width $\Delta=2m$ and jumps to higher values than the gapless
conductivity at $\omega=2m$. For the asymmetric valleys the conductivity behaves similarly,
with the conductivity reduced by a factor $1/2$ inside the gap though.

{\it Hall conductivity}: The form of $h_3$ in Eq. (\ref{h3}) allows us to change 
the gap parameter at the two
valleys separately. This means that the Hall conductivities of the two valleys either
substract each other (for ${\rm sgn} (h_{3,1})={\rm sgn} (h_{3,2})$) or add each other 
(for ${\rm sgn} (h_{3,1})=-{\rm sgn} (h_{3,2})$), when $h_{3,j}$ is the gap parameter at valley $j$
(cf. Eq. (\ref{fullHall})).
The reason is that the Hall conductivities can be evaluated at each gapped valley 
separately, using the low-energy result of Eq. (\ref{eq:hallcond1}). Thus, for the
asymmetric valleys, where one valley is gapped and the other is gapless, we have only
a contribution of $e^2{\rm sgn}(m)/2h$ from the gapped valley. This is what we see 
at low frequencies in Fig. \ref{fig:Hall}. Moreover, there is a logarithmic singularity
at $\omega=2m$. It is not related to the van Hove singularity but appears when the 
frequency of the external field reaches the gap energy. Apparently, the optical Hall
conductivity increases dramatically as the states at the edge of the gap start
to contribute to transport \cite{hill10a}. The position of this singularity is quite
robust and only determined by the gap width, in contrast to the parameter dependent
position of the van Hove singularity. In particular, the properties of the optical
Hall condutivity do not change over a wide range of temperatures (cf. Fig. \ref{fig:Hall}).
This remarkable effect can be used to measure the gap within a transport measurement.
This offers an alternative to other methods of measuring spectral properties through 
transport properties \cite{zhang08}.

In this work we have suggested a possible valley symmetry breaking by a periodic magnetic
flux. The latter is generated by doping of the graphene sheet with spin 1/2 atoms.
The periodic flux opens gaps at both valleys whose values can be controlled independently.
However, in contrast to a homogeneous magnetic field, it does not create Landau levels.
Consequently, at most two Hall plateaux can be observed. Our calculations have revealed that
a gap $\Delta=2|m|$ in both valleys creates the usual Hall plateaux with 
$\sigma_{12}=e^2/h$ when the gap parameter has opposite signs at the two
valleys but a vanishing Hall conductivity if the signs of $m$ are equal. The Hall
conductivity is $\sigma_{12}=e^2/2h$ if one valley is gapped and the other is gapless.
Moreover, the optical Hall conductivity has a logarithmic singularity when the frequency
of the external microwave field reaches the gap energy. This singularity is very robust
and should be visible even at room temperature. Therefore, it can be used for an accurate
determination of the gap by measuring the optical properties of graphene.

\section*{ACKNOWLEGEMENTS}

We acknowledge financial support by the DPG-grant ZI 305/5-1.

\end{document}